# The Elusive member of the Ti-Al-C MAX family- $Ti_4AlC_3$

Subhajit Sarkar, Pratim Banerjee and Molly De Raychaudhury[*]

Department of Physics, West Bengal State University, Berunanpukuria, North 24 PGNS, Kolkata – 700126, INDIA

[*]Corresponding Author Email: molly@wbsu.ac.in

We report here perhaps the first successful synthesis and structural characterization of the n=3 family member of $Ti_{n+1}AlC_n$, i. e. $Ti_4AlC_3$. X-ray Powder diffraction (XRD) data shows characteristic reflections of from corresponding to reflections from the (002), (004), (006), (008), (100), (102), (104), (0010), (105), (106), (0012), (1011) and (1012) planes at $2\theta = 7.64^0$, $15.17^0$, $22.76^0$, $30.5^0$, $35^0$, $37.4^0$ $38.3^0$, $39.2^0$, $41.3^0$, $46.22^0$, $55.24^0$, $58.62^0$ and $60.78^0$ (double structure) respectively. Rietveld refinement of the XRD data reveals a phase purity of about 79 % for $\alpha-Ti_4AlC_3$, 15 % for $\beta-Ti_4AlC_3$ and the rest mostly that of cubic TiC (6 %). The primary crystal symmetry of the two dominant phases is the hexagonal $P6_3/mmc$. The precursors chosen were $TiH_2$, Al metal powder and Carbon powder in a molar ratio of 3:1.2:2, which build the case for an Al-deficient condition. We adopted the pressureless sintering technique at $1350^0$ C with a dwelling time of 4 hours under ultra-high vacuum of $10^{-7}$ mbar. The co-existence of trace amount of $Ti_2AlC$ at $1350^0$ C is proven by the small structure at $2\theta=13.13^0$. No trace of oxides like $Al_2O_3$ or $TiO_2$ was found in the end product. The line profile width of XRD data indicates average grain size of the order of micro meter. The Scanning Electron Microscopy images show highly lamellar stacked growth of almost a pure MAX ($\alpha$ or $\beta$) phase and grain size of micron order, agreeing well with the XRD data.

Among the 150 different MAX compositions discovered in the last 60 years, the most extensively studied ones are the Ti-Al-C and Ti-Al-N systems. And among these two families of $Ti_{n+1}AlC_n$ and $Ti_{n+1}AlN_n$, the most straight forward to synthesize are the n=1 systems in both the families till date. It is worth noting here that $Ti_3AlC_2$ [1-2] was successfully synthesized almost three decades after $Ti_2AlC$ [3]. Although both exhibit excellent ceramic features, metallic property [1-7], stability and machinability properties, their synthesis technique is yet to be fully standardized. The possibility of synthesis of an even higher family member in the series, namely $Ti_4AlC_3$, and investigating its physical properties are worth pursuing. While the two most elusive members are $Ti_3AlN_2$ and $Ti_4AlC_3$, the first ever 413 member materialized in $Ti_4AlN_3$ [8-11], six years after the discovery of the 312 phase of Ti-Al-C phase. Taking into consideration all the reported work till date on the Ti-Al-C MAX phases, we find $Ti_4AlC_3$ has not yet been synthesized, either in pure phase or otherwise. We report below perhaps the first ever synthesis of $Ti_4AlC_3$ and its structural characterization.

The process of synthesis of MAX, aimed at obtaining it in its purest phase, has undergone a lot of iterations over the years. One of the most widely accepted routes to MAX involves ball-milling of the precursors, cold-pressing and sintering at a temperature (ranging between $1300^0$ C and $1650^0$ C) for a duration of 4-24 hours in Ar or vacuum condition [2, 7-12]. Formation of (211) Ti-Al-C phase was reported at a relatively lower temperature of about $1100^0$C. However there are non-trivial issues which prohibit the formulation of a generic process for the synthesis of the family members of a MAX phase. These issues are, namely, the choice of the initial chemicals (precursors), the molar ratio of these ingredients, application of hot isostatic pressure during sintering and the sintering temperature.

The precursors chosen for this particular work are $TiH_2$ (99 % pure , 325 Mesh powder, Alfa Aesar, USA), Al (metal powder, 325 mesh, SRL, India) and Carbon powder (99+ % pure, STREM CHEMICALS INC., USA) in 3:1.2:2 molar ratio. These precursors had been earlier chosen for the synthesis of $Ti_3AlC_2$ [13,14]. Hence an amount of 1.452 g of $TiH_2$ was mixed with 0.3145 g of Al metal powder and 0.233 g of Carbon powder to yield 2 g of the mixture accordingly. This mixture was milled in a polycarbonate vial for 100 minutes with zirconia balls in a ball to powder weight ratio of 3.2143:1 in a SPEX 8000 Mixture Mill, USA. Then 200 mg of this milled mixture was pelletized by isostatically cold-pressing at 7 tons (750 KPa) using a Hydraulic Pellet Press of Kimaya Engineers, India. The pellet was of 10 mm diameter. The weight of the pellet remained very close to 200mg and the pellet was intense black in colour. This pellet was then sintered in a box furnace of Lakshmi Vacuum Technologies, India, under ultra-high vacuum of $10^{-7}$ mbar at $1350^0$ C with appropriate heating rate and a dwelling time of 4 hours. It is to be noted that sintering did not include application of hot isostatic pressure. Although it is claimed [13] that MAX synthesis by Spark Plasma sintering yields purest phase of MAX, recently pressureless sintering[15] has also yielded $Ti_3AlC_2$ of very high purity but at $1400^0$ C. In this case, sintering was followed by appropriate cooling with the vacuum condition maintained till $260^0$ C. The end product (called S hereafter) turned silverish off-white on the surface with a weight loss of about 20 mg.

The crystal structure of this silverish off-white coloured pellet (polycrystalline sample) was investigated using a RIGAKU powder X-ray diffractometer (bench top) with Cu $K_α$ X-ray source of wavelength 1.5406 $A^0$. Fig. S1 (in Supplementary information file) shows the XRD for as-sintered pellet. Compared with the XRD pattern and cif file (No. 9012564) of cubic TiC, we find the XRD in fig. S1 to be dominated by the cubic phase of TiC with peaks at $2θ=36.14^0, 42^0$ and $61.3^0$ arising out of reflections from (111), (200) and (220) set of planes respectively. There was no trace of any structure/peak, in the XRD spectrum, characteristic of any MAX of type $Ti_{n+1}AlC_n$. A portion of this pellet was then pulverized. It is worth mentioning here that the pellet was not at all brittle and it took some effort to be pulverized. The structural characterization of the pulverized sample (called PS hereafter) was done using the same diffractometer. Fig. 1 is the XRD spectrum for the sample PS. Structures and peaks are observed at $2\theta = 7.64^0$, $13.13^0$, $15.17^0$, $22.76^0$, $30.5^0$, around $34^0$-$35^0$, $36.14^0$, $37.4^0$, $38.3^0$, $39.2^0$, $40.9^0$, $41.3^0$, $42^0$, $43.6$, $45.69^0$, $46.22^0$, $55.24^0$, $58.62^0$, $60.78^0$, $62.07^0$, $65.75$, $69.56$ and around $72.7^0$ and $76.5^0$. Comparing with the XRD peaks reported [10] for $Ti_4AlN_3$ (possibly in the $\alpha$ phase of the 413-MAX) and $\alpha$-phase [16] of $Ta_4AlC_3$, we find matching at $2\theta = 7.64^0$, $15.17^0$, $22.76^0$, $30.5^0$, $35^0$, $37.4^0$ $38.3^0$, $39.2^0$, $41.3^0$, $46.22^0$, $55.24^0$, $58.62^0$ and $60.78^0$ corresponding to reflections from the (002), (004), (006), (008), (102), (104), (0010), (105), (106), (0012), (1011) and (1012) planes respectively of the hexagonal type crystal lattice with space group $P6_3/mmc$. All these structures except for the ones at $37.4^0$, $39.2^0$, $41.3^0$ and $60.78^0$ are also found to be compatible with the XRD pattern of $\beta$-phase[16] of $Ta_4AlC_3$. Other significant XRD structures for the sample PS around $2\theta = 34^0$, $40.9^0$, $42^0$ and $62.07^0$ tally with the corresponding ones in the $\beta$-phase[16] of $Ta_4AlC_3$ only and hence can be attributed to $\beta$-phase of $Ti_4AlC_3$. It is also observed that the peaks at $2\theta = 36.14^0$, $42^0$ and $62.07^0$ correspond to the reflections from (111), (200) and (220) planes of the cubic TiC respectively. The insignificant presence of the 211 phase is indicated by a very weak line at $2\theta = 13.13^0$. The Rietveld analysis is performed using Maud software [17] and the lines fitted with Gaussian profile to cubic TiC (cif file no. 9012564), $Ti_4AlN_3$ (cif file no. 1526338) and a modified file of the latter to mimic $\beta$ phase, leads to a goodness of fit of about 4.1 when the peak at $7.64^0$ is considered to be the fundamental one for $\alpha$-$Ti_4AlC_3$ and the one at $38.3^0$ is considered to be the fundamental one for $\beta$-$Ti_4AlC_3$ and that at $42^0$ as the fundamental peak for cubic TiC. The lattice parameters for $\alpha$-$Ti_4AlC_3$ after Rietveld fitting are $a=3.15$ $A^0$ and $c=23.67$ $A^0$. The same for $\beta$-$Ti_4AlC_3$ after Rietveld fitting are $a=3.06$ $A^0$ and $c=23.62$ $A^0$ and for cubic TiC is $a=4.12$ $A^0$. The corresponding atomic positions in the unit cell of space group $P6_3/mmc$ are given in Table I. The weight percent proportion reveals a phase purity of about 79 % for $\alpha$–$Ti_4AlC_3$, 15 % for $\beta$–$Ti_4AlC_3$ and the rest mostly that of cubic TiC (6%) with less than 1% for $Ti_2AlC$. It is noteworthy that all reports on successful synthesis of M-Al-C MAX phases (M=Ti, Ta) mention coexistence of cubic TiC phase and stability of $Ti_2AlC$ at least till $1350^0$ C. Pure $Ti_3AlC_2$ was claimed to be successfully synthesized by both Barsoum et al. [2] and Desai et al. [13] but at sintering temperatures of $1350^0$ C and $1400^0$ C respectively. In the light of this information, first ever synthesis of 94% phase proportion of $Ti_4AlC_3$ (in $\alpha$ and $\beta$ phases) at $1350^0$ C is indeed very promising for this class of materials. This also holds significance for the minimum sintering temperature required for the synthesis of $Ti_4AlC_3$ and possibly for other carbide MAXs. Here we would like to emphasize on a crucial issue of the molar ratio in which the precursors are taken for the synthesis of the desired MAX. The issue of proportion between Al and C is more crucial than the proportion between $TiH_2$ and Al. The greater the proportion of Al, the more is the possibility of formation of the lower family member of the $Ti_{n+1}AlC_n$ series. This shall be dealt with in more details in a forthcoming report soon.

The Fourier-transformed Infra-red spectrum for the polycrystalline pulverized sample (PS) was measured at room temperature using a FTIR/FIR Spectrometer (Frontier

Model) of PerkinElmer make. Fig. 2 depicts the FTIR spectrum with major absorption structures at 414 cm$^{-1}$, 474 cm$^{-1}$, 601 cm$^{-1}$, 995 cm$^{-1}$ and 1113 cm$^{-1}$. Measurements were made starting from 300 cm$^{-1}$ and higher. It is well-known that FTIR spectroscopy gives the structures corresponding to absorption or transmittance only due to the vibration of heteronuclear bonds. Ti$_4$AlN$_3$ is known to have a total of 10 Raman-active modes out of which three are also infrared-active.[18] These are the E$_{1g}$ modes occurring at 132 cm$^{-1}$, 220 cm$^{-1}$ and 588 cm$^{-1}$ corresponding to vibrations involving Ti-Al atoms and Ti-N atoms.[19] Hence it can be analogically concluded that the peak at 601 cm$^{-1}$ in Ti$_4$AlC$_3$ is certainly one of those E$_{1g}$ modes, involving atoms in Ti1, Ti2 and C2 layers. The symbol '1' stands for inner layers and '2' stands for outer layers. The lower two infrared-active modes found at 414 cm$^{-1}$ and 474 cm$^{-1}$ should involve Ti-Al vibrations and the two higher ones at 995 cm$^{-1}$ and 1113 cm$^{-1}$ should mainly involve Carbon atoms. The characteristic bondlengths of the major α- phase of the sample PS extracted from visualization software VESTA [20] for Ti1-C1, Ti1-C2, Ti2-C2, Ti2-Al are 2.207 A$^0$, 2.187 A$^0$, 2.02 A$^0$ and 2.91 A$^0$. Since the Ti-C bonds range from 2.02 A$^0$ to 2.207 A$^0$ which coincides with the corresponding values for Ti-N bonds in Ti$_4$AlN$_3$, the fact that the E$_{1g}$ mode corresponding to Ti-C(N) vibration occurs at almost same value (588 cm$^{-1}$ for Ti$_4$AlN$_3$ and 601 cm$^{-1}$ for Ti$_4$AlC$_3$) is corroborating.

The scanning electron microscopy SEM analysis was performed using aEVO-18 Special Edition microscope of Carl Zeiss make with 15 kV beam, 7.51/7.78/7.68 mm working distance and SE1 detector. No coating was used for observation. Fig. 3 (a) and 3 (b) are the SEM images of the unpulverized portion of the sample S with magnifications of 10000 and 2000 respectively. Both reveal nicely stacked well resolved lamellar structure of (413) MAX phase with smooth surfaces and well-defined grains. Fig. S2 (a) and S2 (b) (in Supplementary information file) are the corresponding SEM images of another area of the pellet. And Fig. S3 (a) and S3 (b) (in Supplementary information file) are the corresponding SEM images of another area of the pellet which clearly shows well-demarcated surface dominated by phase 3, i. e. TiC and sub-surface lamellar structures of α− and β− phase of Ti$_4$AlC$_3$ dominating the almost entire thickness of the pellet. Similar phase analysis [21] using SEM images of Ti$_4$AlN$_3$ yielded coexistence of 4 phases, namely Ti$_4$AlN$_3$, TiN, TiAl$_3$ and Al$_2$O$_3$. It is to be noted that though formation of Ti$_4$AlN$_3$ began at 1275$^0$ C but a 'single' phase was claimed to have been formed upon sintering in that particular in Ar atmosphere with chamber under hot isostatic pressure for 168 hours at 1598 K. In comparison, this work reports a fairly less complicated route at considerably lower sintering temperature than that reported for Ti$_4$AlN$_3$ and with no trace of oxides in the end product. We believe that the ultra-high vacuum during sintering led to the complete absence of Al$_2$O$_3$ and TiO$_2$ in the end product. The former is known to form a protective layer around the lower MAX phase, namely the 211 phase which is formed at a lower temperature [22]. In order to form the higher family series, it is necessary to break the protective Alumina layer around the preliminary MAX and that considerably raises the activation barrier. Hence the 413 phase has been found to be elusive till now. The formation temperature has been lowered by raising the vacuum level during sintering.

Preliminary dc electrical measurements on the pellet of (PS) sample showed that they were conducting. Detailed optical and electrical measurements and data analysis are in progress. Fig. S1, S2 and S3 are provided in Supplementary information.

Table I: Crystal Structure data after Rietveld refinement of the XRD data

| Phase | Lattice parameter a (A$^0$) | Lattice parameter c (A$^0$) | Atom | Site | x | y | z | Occupancy |
|---|---|---|---|---|---|---|---|---|
| α–Ti$_4$AlC$_3$ P6$_3$/mmc | 3.06 | 23.61 | Ti1 | 4f | 1/3 | 2/3 | 0.056 | 1.0 |
| | | | Ti2 | 4e | 0 | 0 | 0.1520 | 1.0 |
| | | | Al | 2c | 1/3 | 2/3 | 1/4 | 1.0 |
| | | | C1 | 2a | 0 | 0 | 0 | 1.0 |
| | | | C2 | 4f | 2/3 | 1/3 | 0.1106 | 1.0 |
| β-Ti$_4$AlC$_3$ P6$_3$/mmc | 3.15 | 23.65 | Ti1 | 4f | 1/3 | 2/3 | 0.0546 | 1.0 |
| | | | Ti2 | 4e | 2/3 | 1/3 | 0.1444 | 1.0 |
| | | | Al | 2c | 1/3 | 2/3 | 1/4 | 1.0 |
| | | | C1 | 2a | 0 | 0 | 0 | 1.0 |
| | | | C2 | 4f | 0 | 0 | 0.1088 | 1.0 |

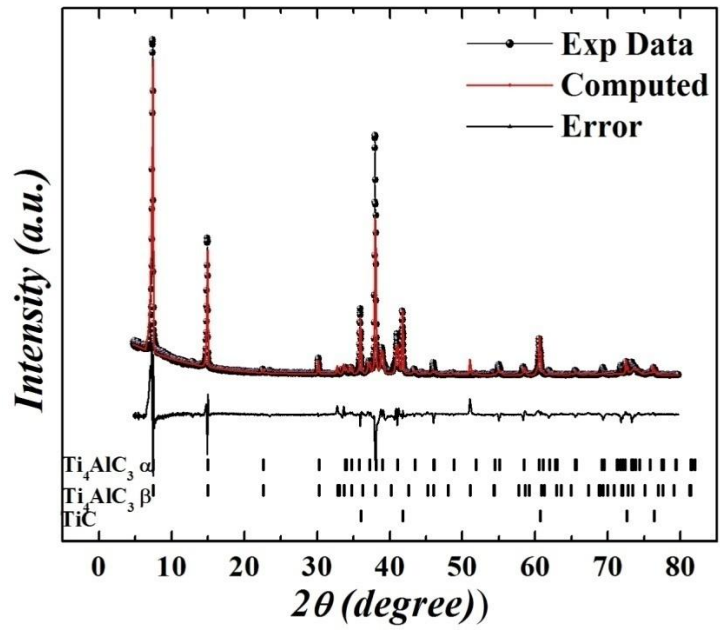

Fig. 1: XRD of pulverized sintered pellet (PS)

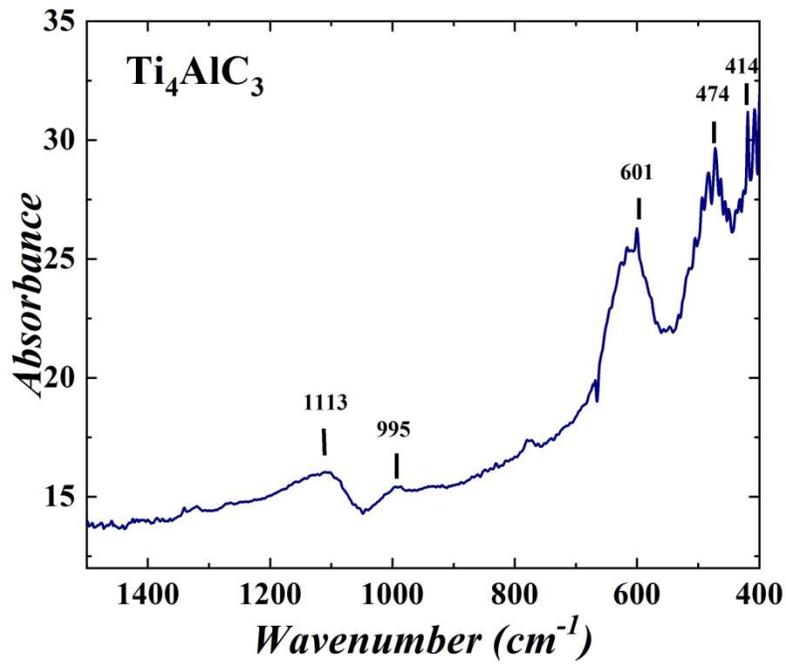

Fig. 2: FTIR of pulverized sintered pellet (PS)

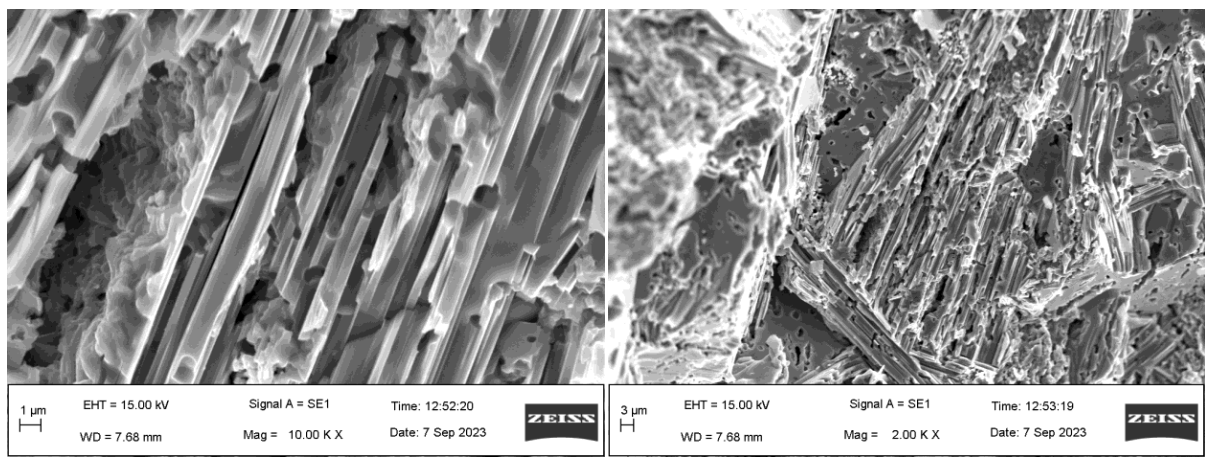

Fig. 3: SEM image of (a) pellet (S) with magnification 10000 X and (b) pellet (S) with magnification 2000 X

**Conclusions**

To the best of our knowledge, this is the first report of a successful synthesis of oxide free-$Ti_4AlC_3$. The other highlight is that this has been achieved at $1350^0$ C, substantially lower than that reported for $Ti_4AlN_3$. The formation of $Ti_4AlC_3$ has been confirmed through XRD, SEM and FTIR studies. The lower formation temperature has been achieved through increasing the vacuum level which in turn did not allow the formation of $Al_2O_3$ and $TiO_2$. Alumina, once formed around the 211 phase, is difficult to be got rid of, thereby enhancing the activation barrier for the formation of the higher family series.

**Acknowledgements**

MDR and PB would like to thank the WBDST, Govt. Of WB, India for financial support vide Project No. 462(Sanc.)/ST/P/S&T/16G-9/2018. Authors would like to acknowledge characterisation support from Mr. Debabrata Samanta, School of Physical Sciences, IISER Kolkata. We would like to extend our sincere appreciation to CRTDH-CSIR-CMERI Durgapur, Govt. of India, for their invaluable contribution in "Heat treatment of the powder samples under running vacuum condition". Their expertise and dedication played a pivotal role in achieving the